\documentclass{appolb}
\usepackage{graphicx}

\usepackage{amsmath, empheq, amsfonts, amssymb}

\begin{document}
\title{Production of $f_0(500)$, $f_0(980)$ and $a_0(980)$ in the $\chi_{c1} \to \eta \pi^+ \pi^-$ and $\eta_c \to \eta \pi^+ \pi^-$ decays%
\thanks{Presented by Vin\'{i}cius Rodrigues Debastiani at Excited QCD - May 2017 - Sintra, Portugal}%
}

\author{V.~R.~Debastiani$^\dag$, Wei-Hong~Liang$^\ddag$, Ju-Jun~Xie$^\S$ and E.~Oset$^\dag$
\address{$^\dag$Departamento de
F\'{\i}sica Te\'orica and IFIC, Centro Mixto Universidad de
Valencia-CSIC Institutos de Investigaci\'on de Paterna, Aptdo.
22085, 46071 Valencia, Spain\\
$^\ddag$Department of Physics, Guangxi Normal University,
Guilin 541004, China
$^\S$Institute of Modern Physics, Chinese Academy of
Sciences, Lanzhou 730000, China
}
\\
}
\maketitle
\begin{abstract}
Using the chiral unitary approach in coupled channels and $SU(3)$ symmetry we describe the production of $f_0(500)$, $f_0(980)$ and $a_0(980)$ in the $\chi_{c1} \to \eta \pi^+ \pi^-$ reaction, recently performed by the BESIII collaboration. A very strong peak for the $a_0(980)$ can be seen in the $\eta\pi$ invariant mass, while clear signals for the $f_0(500)$ and $f_0(980)$ appear in the one of $\pi^+\pi^-$. Next, we make predictions for the analogous decay $\eta_c \to \eta \pi^+ \pi^-$, which could also be measured experimentally. We discuss the differences of these reactions which are interesting to test the picture where these scalar mesons are dynamically generated from the interaction of pairs of pseudoscalars.
\end{abstract}

\section{Introduction}

 The experiment on the $\chi_{c1} \to \eta \pi^+ \pi^-$ decay performed with high statistics by the BESIII collaboration \cite{BES}, and previously by the CLEO collaboration \cite{CLEO}, presents an interesting opportunity to test the picture where the scalar mesons $f_0(500)$, $f_0(980)$ and $a_0(980)$ are dynamically generated from the final state interaction of meson pairs $\pi^+\pi^-$ and $\eta\pi^\pm$. Indeed, it is found that the most dominant two-body structure comes from $a_0(980)^\pm \pi^\mp$,  with $a_0(980)^\pm \to \eta \pi^\pm$.

In this short paper we will briefly discuss the work of Refs. \cite{chic1,etac} where the chiral unitary approach and $SU(3)$ symmetry were used to describe the production of these three scalars in the BESIII experiment and to make predictions for the analogous reaction with $\eta_c$ instead of $\chi_{c1}$. We will make a short discussion on $SU(3)$ scalars and compare the treatment of the amplitude and mass distribution used to describe each decay.

\section{Common Formalism}

We start by considering that the charmonium states $c\bar c$ behave as a $SU(3)$ scalar, and use the following $\phi$ matrix to get the weight of every trio of pseudoscalar mesons created in the $\chi_{c1}$ or $\eta_c$ decay
\begin{equation}
\label{Mphi}
\phi \equiv \left(
           \begin{array}{ccc}
             \frac{1}{\sqrt{2}}\pi^0 + \frac{1}{\sqrt{3}}\eta + \frac{1}{\sqrt{6}}\eta' & \pi^+ & K^+ \\
             \pi^- & -\frac{1}{\sqrt{2}}\pi^0 + \frac{1}{\sqrt{3}}\eta + \frac{1}{\sqrt{6}}\eta' & K^0 \\
            K^- & \bar{K}^0 & -\frac{1}{\sqrt{3}}\eta + \sqrt{\frac{2}{3}}\eta' \\
           \end{array}
         \right).
\end{equation}
If we think of $\phi$ as a $q\bar{q}$ matrix, as discussed in Ref. \cite{chic1}, it is natural to build a $SU(3)$ scalar by taking 
\begin{align}\label{eq:phiphiphi}
\nonumber SU(3){\rm [scalar]} \equiv {\rm Trace} (\phi \phi \phi) &= 2\sqrt{3} \eta \pi^+ \pi^-
   + \sqrt{3} \eta \pi^0 \pi^0
   + \frac{\sqrt{3}}{9} \eta \eta \eta \\
&   + 3 \pi^+ K^0 K^- + 3 \pi^- K^+ \bar K^{0},
\end{align}
where we have already neglected the $\eta'$ which plays only a marginal role in the building of the $f_0(500)$, $f_0(980)$, $a_0(980)$ resonances, because of its large mass and small couplings. We have also neglected the terms that cannot make a transition to the final state $\eta\pi^+\pi^-$.

In fact, there are four $SU(3)$ scalars: ${\rm Trace} (\phi \phi \phi)$, ${\rm Trace}(\phi) {\rm Trace}(\phi \phi)$, $[{\rm Trace}(\phi)]^3$ and ${\rm Det}(\phi)$. But by the Cayley-Hamilton relation,
\begin{equation}
 2 {\rm Trace}(\phi\phi\phi) - 6 {\rm Det}(\phi) - 3 {\rm Trace}(\phi) {\rm Trace}(\phi\phi) + [{\rm Trace}(\phi)]^3 = 0,
\end{equation}
only three of them are independent. In Ref. \cite{etac} we discussed other possibilities and concluded that the best choice is indeed ${\rm Trace} (\phi \phi \phi)$.

Next, we use the chiral unitary approach to describe how the scalar mesons are dynamically generated from the interaction of pairs of pseudoscalars in coupled channels.
We follow the framework of Ref. \cite{Oller}, using an effective chiral Lagrangian where mesons are the degrees of freedom
\begin{eqnarray}
\label{lagran}
\mathcal{L}_2=\frac{1}{12\,f_{\pi}^2}\mathrm{Trace}[\,\, (\partial_{\mu}\phi\,\phi
- \phi\,\partial_{\mu}\phi)^2+M\phi^4\,\,]\, ,
\end{eqnarray}
where $\phi$ is the matrix in Eq. \eqref{Mphi}, $f_\pi$ is pion decay constant and
\begin{equation}
   \quad
     M=
    \left(
    \begin{array}{@{\,}ccc@{\,}}
     m^2_{\pi} & 0 & 0\\
    0 & m^2_{\pi} & 0\\
    0 & 0 & 2 m^2_{K}-m^2_{\pi}\\
  \end{array}
   \right) \, .
\end{equation}

From this Lagrangian we extract the kernel of each channel, which in charge basis are: 1) $\pi^+\pi^-$, 2) $\pi^0\pi^0$, 3) $K^+K^-$, 4) $K^0\bar K^0$, 5) $\eta\eta$, 6) $\pi^0\eta$ and can be found in Refs. \cite{Liang,Xie}. These kernels are used to build the $V$ matrix which is then inserted into the Bethe-Salpeter equation, summing the contribution of every meson-meson loop.
\begin{equation}\label{bs}
T=(1-VG)^{-1}\,V\, ,
\end{equation}
where $G$ is the meson-meson loop function, which we regularize with a cutoff using $q_{\rm max}\sim600$ MeV. After the integration in $q^0$ and $\cos\theta$ we have
\begin{equation}
G=\int_0^{q_{\rm max}}\frac{q^2dq}{(2\pi)^2}\frac{\omega_1 + \omega_2}{\omega_1\omega_2[(P^0)^2-(\omega_1 + \omega_2)+i\epsilon]}\ ,
\label{eq:loopex}
\end{equation}
with $\omega_i = \sqrt{q^2+m_i^2}$, $P^0 = s$. Each kernel is projected in $S$-wave and a normalization factor is included when identical particles are present, which later needs to be restored. Finally, the $T$ matrix will give us the scattering and transition amplitudes between each channel, and isospin symmetry is used to obtain the amplitude of channels with different charges \cite{chic1}.

\section{Theoretical description of $\chi_{c1} \to \eta \pi^+ \pi^-$}

Following the assumption that $c\bar{c}$ behaves as a $SU(3)$ scalar, we look at the quantum numbers of the initial and final states, combining them in two cases: $\eta$ leaves in $P$-wave while $\pi^+\pi^-$ go through final state interaction with $I=0$ to form the $f_0(500)$ and $f_0(980)$ in $S$-wave; and $\pi^-$ (or $\pi^+$) leaves in $P$-wave while $\eta\pi^+$ (or $\eta\pi^-$) go through final state interaction with $I=1$ to form the $a_0^{\pm}(980)$ in $S$-wave.
\begin{figure}[htb]
\centerline{%
\includegraphics[width=12.5cm]{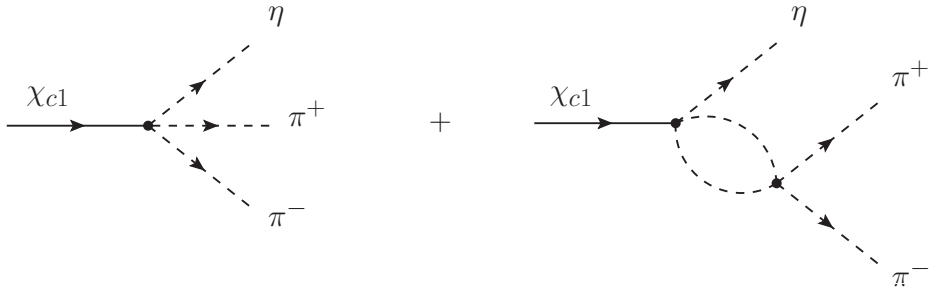}}
\caption{Diagrams considered in the description of $f_0(500)$ and $f_0(980)$ production in $\chi_{c1} \to \eta \pi^+ \pi^-$ reaction: tree-level (left) and reescatering of $\pi^+ \pi^-$ pair (right).}
\label{Fig:chic1diagsum}
\end{figure}

To illustrate our method, we will describe the case where $\eta$ leaves in $P$-wave and $\pi^+\pi^-$ interact. In this case we will consider the diagrams of Fig. \ref{Fig:chic1diagsum}. Then from the $SU(3)$ scalar in Eq. \eqref{eq:phiphiphi}, we select the terms in which we can isolate one $\eta$ and let the other pairs reescater, since our coupled channels approach allows them to make a transition to $\pi^+\pi^-$ final state,
\begin{equation}\label{eq:eta_term}
\eta \left( 2\sqrt{3} \pi^+ \pi^-
+ \sqrt{3} \pi^0 \pi^0 + \frac{\sqrt{3}}{9} \eta \eta \right).
\end{equation}

Then we will have the sum of tree-level and reescatering:
\begin{equation}\label{eq:t_eta}
t_{\eta}= V_P \left( \vec{\epsilon}_{\chi_{c1}} \cdot \vec{p}_{\eta} \right)  \left( h_{\pi^+ \pi^-}+\sum_i h_i S_i G_i[M_{\rm inv}(\pi^+ \pi^-)] t_{i,\pi^+ \pi^-}[M_{\rm inv}(\pi^+ \pi^-)] \right),
\end{equation}
where $h_i$ are the weights of Eq. \eqref{eq:eta_term}, $S_i$ are symmetry and combination factors for the identical particles and the factor $V_P$ provides a global normalization, which is fitted to the data in the $a_0(980)$ peak.

Finally, we can write the differential mass distribution for $\pi^+ \pi^-$
\begin{equation}\label{eq:dGamma_pipi}
  \frac{d \Gamma}{d M_{\rm inv}(\pi^+\pi^-)}
  =\frac{1}{(2\pi)^3}\frac{1}{4M_{\chi_{c1}}^2}\frac{1}{3}p_{\eta}^2 p_{\eta} \tilde{p}_{\pi} \left| t_{\eta} \right|^2,
\end{equation}
where $p_{\eta}$ is the $\eta$ momentum in the $\chi_{c1}$ rest frame and $\tilde{p}_{\pi}$ is the pion momentum in the $\pi^+ \pi^-$ rest frame.

\begin{figure}[htb]
\centerline{%
\includegraphics[width=6.7cm]{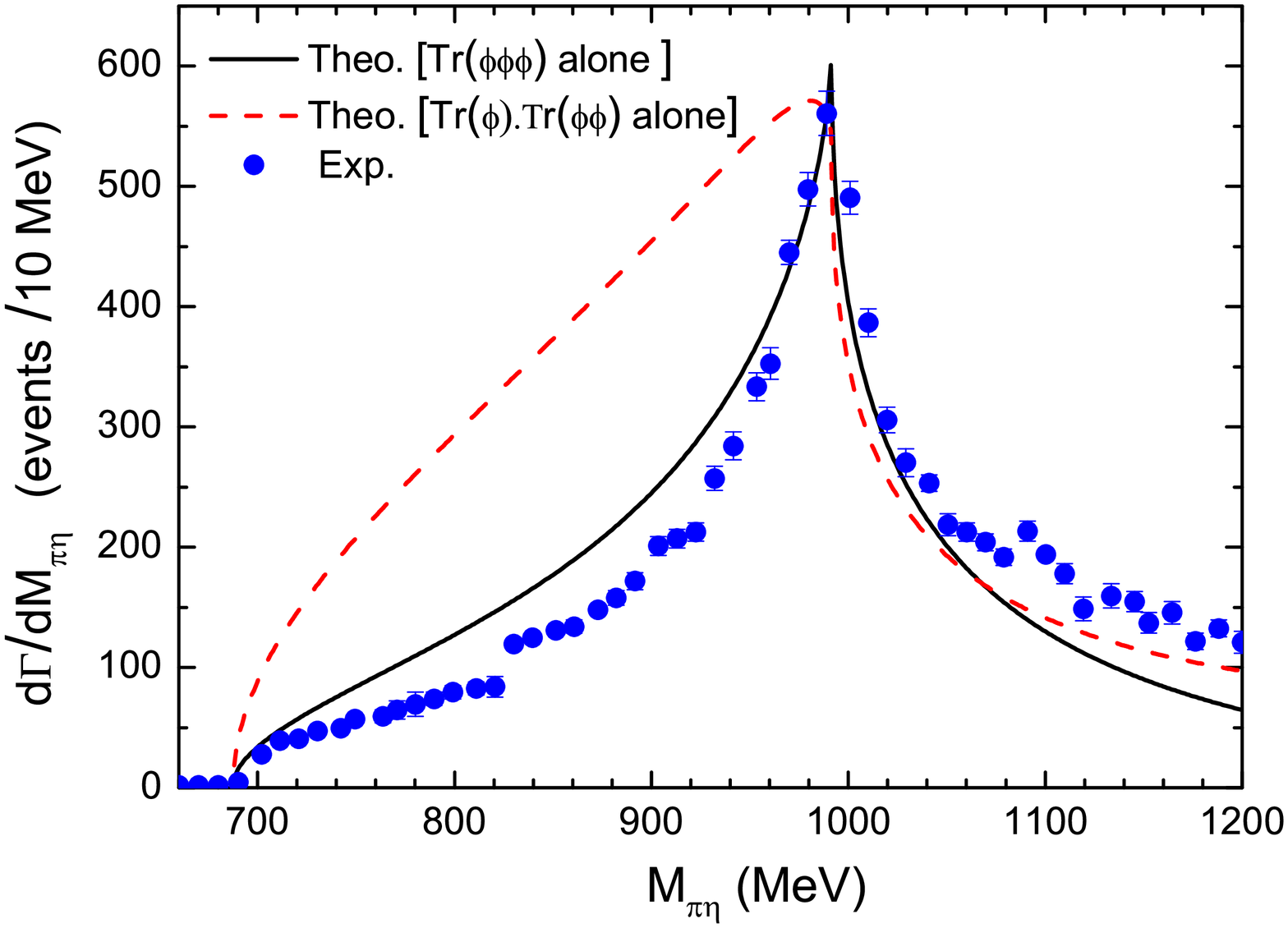}
\includegraphics[width=6.7cm]{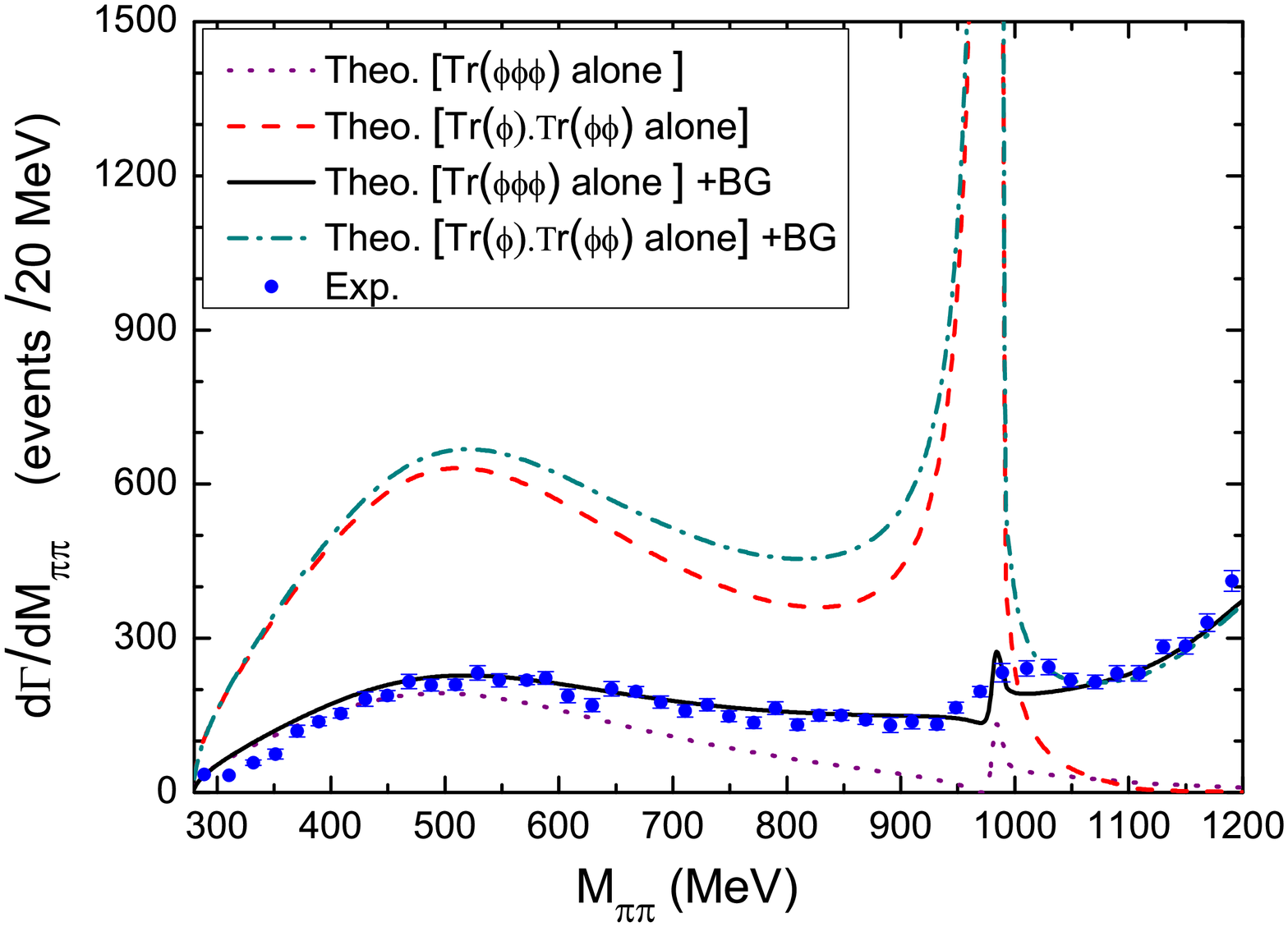}}
\caption{Results for the $\pi\eta$ (left) and $\pi^+ \pi^-$ (right) mass distribution in the $\chi_{c1} \to \eta \pi^+ \pi^-$ reaction, using ${\rm Trace}(\phi\phi\phi)$ or ${\rm Trace}(\phi){\rm Trace}(\phi\phi)$. A linear background is fitted to the data in the $\pi^+ \pi^-$ mass distribution. Data from Ref. \cite{BES}.}
\label{Fig:chic1Trace}
\end{figure}

In Fig. \ref{Fig:chic1Trace} we show the results using the method of Ref. \cite{chic1} and the experimental data of Ref. \cite{BES}. We also compare the results using ${\rm Trace}(\phi\phi\phi)$ as the $SU(3)$ scalar to the case where only ${\rm Trace}(\phi){\rm Trace}(\phi\phi)$ was used, and see that the later is completely off from experiment.

\section{Predictions for $\eta_c \to \eta \pi^+ \pi^-$}

In the analogous reaction $\eta_c \to \eta \pi^+ \pi^-$ the dominant structure will be the one where every final state meson goes out in $S$-wave. Therefore one must consider the interference between each term in the amplitude, then
\begin{equation}
t = t_{tree} + t_{\eta} + t_{\pi^+} + t_{\pi^-}, \qquad t_{tree} = V_P\, h_{\eta \pi^+ \pi^-}.
\end{equation}

Each of the later three terms is a function of an invariant mass, analogous to Eq. \eqref{eq:t_eta}. We select $M_{\rm inv}(\pi^+\pi^-)$ and $M_{\rm inv}(\pi^+\eta)$ as variables and the third one is determined by the relation:
$M^2_{13} = M^2_{\eta_c} + 2 m^2_{\pi} + m^2_{\eta} - M^2_{12} - M^2_{23}$.
It is also necessary to consider the double differential mass distribution \cite{PDG}
\begin{align}\label{eq:d2Gamma}
\begin{aligned}
\frac{d^2 \Gamma}{d M_{\rm inv}(\pi^+ \pi^-) d M_{\rm inv}(\pi^+ \eta)}
= \frac{1}{(2\pi)^3} \frac{1}{8 M_{\eta_c}^3} M_{\rm inv}(\pi^+ \pi^-) M_{\rm inv}(\pi^+ \eta) |t|^2,
\end{aligned}
\end{align}
where we need to integrate in one of the invariant masses to get the distribution of the other one. This way the background of $\pi^+ \eta$ appears naturally in the $\pi^+\pi^-$ mass distribution and vice-versa.

Since our approach is valid only for energies up to 1.2 GeV,
we need to introduce a cut in each amplitude to perform the integration.
To do that we evaluate $Gt(M_{\rm inv})$ combinations up to $M_{\rm inv}=M_{\rm cut}$. From there on, we multiply $Gt$ by a smooth factor to make it gradually decrease at large $M_{\rm inv}$,
\begin{equation}\label{Gt}
  Gt(M_{\rm inv}) = Gt(M_{\rm cut}){\rm e}^{-\alpha(M_{\rm inv}-M_{\rm cut})},
  ~~ {\rm for} ~~M_{\rm inv} > M_{\rm cut}.
\end{equation}

\begin{figure}[htb]
\centerline{%
\includegraphics[width=6.5cm]{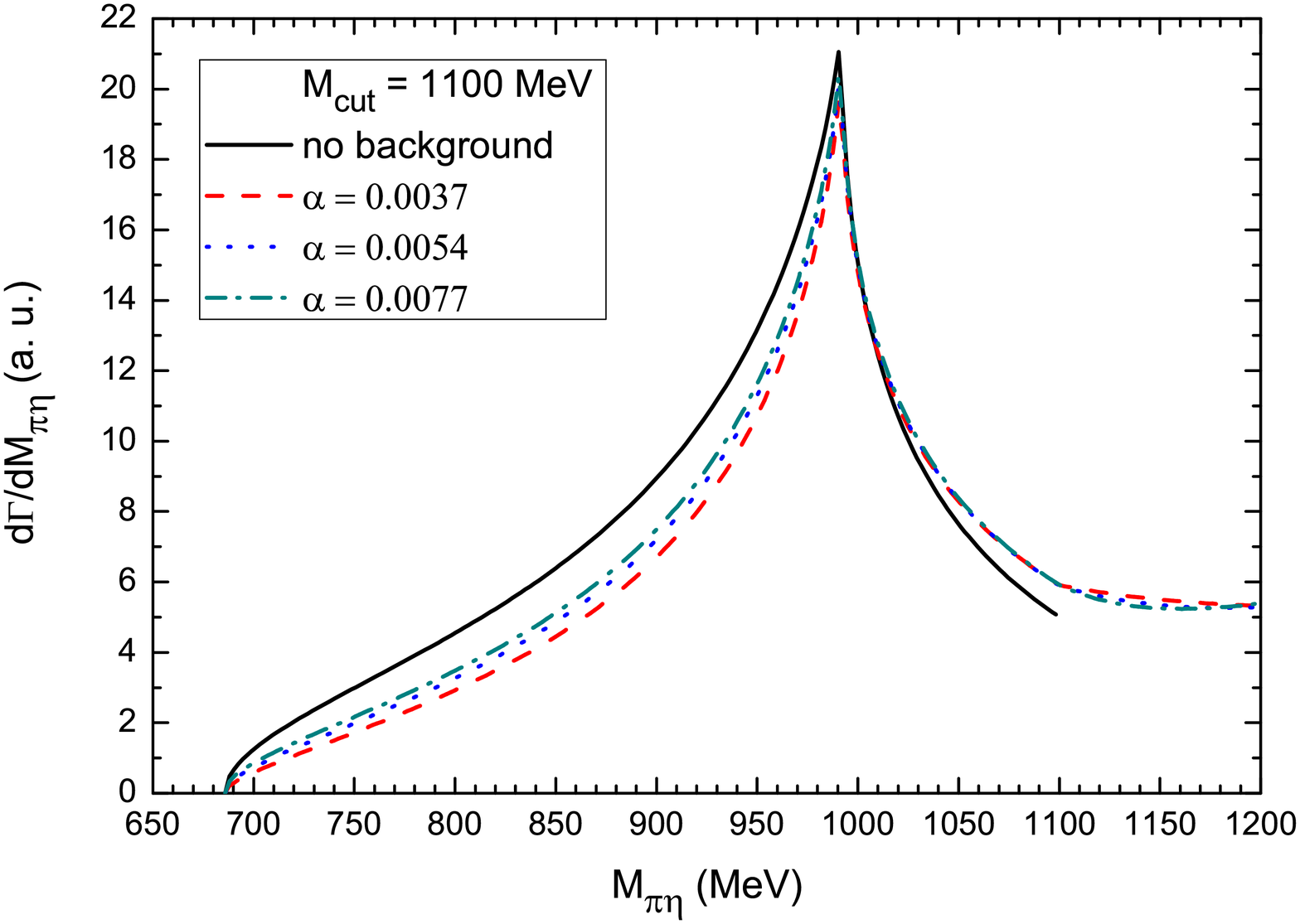}
\includegraphics[width=6.5cm]{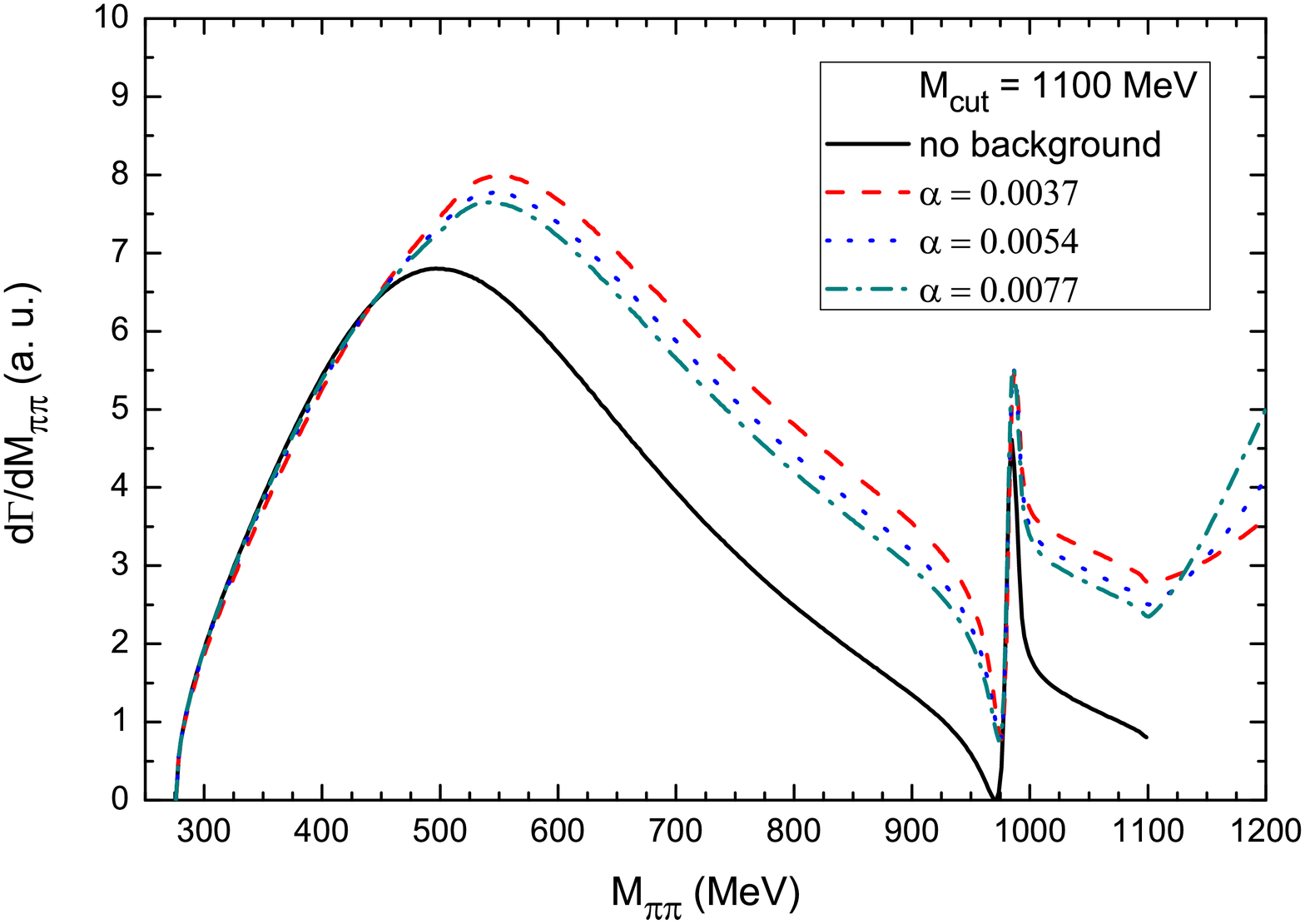}}
\caption{Predictions from Ref. \cite{etac} for the mass distribution of $\pi\eta$ (left) and $\pi^+\pi^-$ (right) in $\eta_c \to \eta \pi^+ \pi^-$, using $M_{\rm cut}=1100$ MeV and $\alpha=$ 0.0037, 0.0054, 0.0077 MeV$^{-1}$, which reduce $Gt$ by a factor 3, 5 and 10, respectively, at $M_{\rm cut} + 300$ MeV. The ``no background'' curve is obtained by keeping only the tree-level and the main reescatering amplitude.}
\label{Fig:etacResults}
\end{figure}

In Fig. \ref{Fig:etacResults} we show the predictions for the production of $f_0(500)$, $f_0(980)$ and $a_0(980)$ in the $\eta_c \to \eta \pi^+ \pi^-$ decay. To compare qualitatively with the results of the previous section, we show with the solid curves, denoted by ``no background'', the results obtained by keeping only the tree-level and the main reescatering amplitude $t_{\pi^-}[M_{\rm inv}(\pi^+\eta)]$ in the case of $a_0(980)$ and $t_{\eta}[M_{\rm inv}(\pi^+\pi^-)]$ in the case of the $f_0(500)$ and $f_0(980)$. We can see that the background introduced goes in the direction where there was a small discrepancy between the results of Ref. \cite{chic1} and the data of Ref. \cite{BES} in the $\chi_{c1} \to \eta \pi^+ \pi^-$ reaction.

\section*{Acknowledgments}

We would like to thank N. Kaiser for information concerning $SU(3)$ invariants. V. R. Debastiani wishes to acknowledge the organizers of the event and the support from the Programa Santiago Grisol\'ia of Generalitat Valenciana (GRISOLIA/2015/005). E. Oset wishes to acknowledge the support from the Chinese Academy of Science in the Program of Visiting Professorship for Senior International Scientists (Grant No. 2013T2J0012).
This work is partly supported by the National Natural Science Foundation of China under Grants No. 11565007, No. 11547307 and No. 11475227. It is also supported by the Youth Innovation Promotion Association CAS (No. 2016367).
This work is also partly supported by the Spanish Ministerio de Economia y Competitividad and European FEDER funds under the contract number FIS2014-57026-REDT, FIS2014-51948-C2-1-P and FIS2014-51948-C2-2-P, and the Generalitat Valenciana in the program Prometeo II-2014/068.

\bibliographystyle{plain}

\begin{thebibliography}{99}

\bibitem{BES} M.~Ablikim {\it et al.} [BESIII Collaboration],
    Phys.\ Rev.\ D {\bf 95}, no. 3, 032002 (2017).

\bibitem{CLEO}
  G.~S.~Adams {\it et al.} [CLEO Collaboration],
  Phys.\ Rev.\ D {\bf 84}, 112009 (2011).

\bibitem{chic1} W.~H.~Liang, J.~J.~Xie and E.~Oset,
     Eur.\ Phys.\ J.\ C {\bf 76}, no. 12, 700 (2016).

\bibitem{etac}
    V.~R.~Debastiani, W.~H.~Liang, J.~J.~Xie and E.~Oset,
     Phys.\ Lett.\ B {\bf 766}, 59 (2017).

\bibitem{Oller}
  J.~A.~Oller and E.~Oset,
  Nucl.\ Phys.\ A {\bf 620}, 438 (1997);
  Erratum: [Nucl.\ Phys.\ A {\bf 652}, 407 (1999)].

\bibitem{Liang}
  W.~H.~Liang and E.~Oset,
  Phys.\ Lett.\ B {\bf 737}, 70 (2014).

\bibitem{Xie}
  J.~J.~Xie, L.~R.~Dai and E.~Oset,
  Phys.\ Lett.\ B {\bf 742}, 363 (2015).

\bibitem{PDG}
  C.~Patrignani {\it et al.} [Particle Data Group Collaboration],
  Chin.\ Phys.\ C {\bf 40}, no. 10, 100001 (2016).

\end{thebibliography}

\end{document}